\documentclass[aps,prc,twocolumn,showpacs,superscriptaddress,footinbib]{revtex4-1}

\usepackage{graphicx}
\usepackage{dcolumn}
\usepackage{bm}
\usepackage{enumerate}

\usepackage[normalem]{ulem}  
\usepackage[dvips]{color}

\begin{document}


\title{Statistical multifragmentation model with discretized energy\\ and the generalized Fermi breakup. \\
I. Formulation of the model}

\author{S.R. Souza}
\affiliation{Instituto de F\'\i sica, Universidade Federal do Rio de Janeiro Cidade Universit\'aria, \\CP 68528, 21941-972, Rio de Janeiro, Brazil}
\affiliation{Instituto de F\'\i sica, Universidade Federal do Rio Grande do Sul,\\
Av. Bento Gon\c calves 9500, CP 15051, 91501-970, Porto Alegre, Brazil}
\author{B.V. Carlson}
\affiliation{Departamento de F\'\i sica, Instituto Tecnol\'ogico de Aeron\'autica-CTA, 12228-900, S\~ao Jos\'e dos Campos, Brazil}
\author{R. Donangelo}
\affiliation{Instituto de F\'\i sica, Universidade Federal do Rio de Janeiro Cidade Universit\'aria, \\CP 68528, 21941-972, Rio de Janeiro, Brazil}
\affiliation{Instituto de F\'\i sica, Facultad de Ingenier\'\i a, Universidad de la Rep\'ublica, Julio Herrera y Reissig 565, 11.300 Montevideo, Uruguay}
\author{W.G. Lynch}
\affiliation{National Superconducting Cyclotron Laboratory and Department of Physics and Astronomy Department,\\ Michigan State University, East Lansing, Michigan 48824, USA}
\author{M.B. Tsang}
\affiliation{National Superconducting Cyclotron Laboratory and Department of Physics and Astronomy Department,\\ Michigan State University, East Lansing, Michigan 48824, USA}

\date{\today}

\begin{abstract}
The Generalized Fermi Breakup recently demonstrated to be formally equivalent to the Statistical Multifragmentation Model, if the contribution of excited states are included in the state densities of the former, is implemented.
Since this treatment requires the application of the Statistical Multifragmentation Model repeatedly on the hot fragments until they have decayed to their ground states, it becomes extremely computational demanding, making its application to the systems of interest extremely difficult.
Based on exact recursion formulae previously developed by Chase and Mekjian to calculate the statistical weights very efficiently, we present an implementation which is efficient enough to allow it to be applied to large systems at high excitation energies.
Comparison with the GEMINI++ sequential decay code shows that the predictions obtained with  our treatment are fairly similar to those obtained with this more traditional model.  
\end{abstract}

\maketitle

\begin{section}{Introduction}
\label{sect:introduction}
The theoretical understanding of many nuclear processes requires the treatment of the de-excitation of the reaction products, since most of them have already decayed (typically within $10^{-20}$~s) by the time they are observed at the detectors (after $10^{-9}$~s).
Owing to the great complexity associated with the theoretical description of the nuclear decay, different approaches have been developed over many decades, ranging from the pioneering fission treatment of Bohr-Wheeler \cite{BohrWheeler} and the Weisskopf-Ewing statistical emission \cite{Weisskopf} to the modern GEMINI binary decay codes \cite{GEMINI1988} and the GEMINI++ \cite{GEMINIpp2010_1,GEMINIpp2010_2}, which generalize the Bohr-Wheeler treatment.
Many other models, which focus on different aspects of the de-excitation process, such as pre-equilibrium emission, have also been developed by other groups (see Ref.\ \cite{ColeBook} for an extensive review on the statistical decay treatments).
 
The decay of complex fragments produced in reactions that lead to relatively hot sources, whose temperature is higher than approximately 4 MeV, has often been described by simpler models, due to the very large number of primary hot fragments produced in different reaction channels, making the need for computational efficiency, at least, as important as the corresponding accuracy.
For this reason, treatments based on the Weisskopf-Ewing decay and the Fermi-Breakup model \cite{FermiBreakup1,FermiBreakup2} have been extensively employed \cite{grandCanonicalBotvina1987,Bondorf1995} in these cases.
More evolved schemes, such as the MSU-decay \cite{ISMMlong,BettyPhysRep2005}, which incorporates much empirical information besides employing the GEMINI code and the Hauser-Feshbach formalism \cite{HauserFeshbach} where such information is not available, have also been developed.

Recently, a generalization of the Fermi-Breakup model (GFBM), including contributions from the density of excited states, has been proposed in Ref.\ \cite{fbk2012} and demonstrated to be formally equivalent to the Statistical Multifragmentation Model (SMM) \cite{smm1,smm2,smm4}.
However, the inclusion of the density of excited states makes the GFBM considerably more computationally involved than its simplified traditional version \cite{grandCanonicalBotvina1987} in which only very few discrete excited states are considered.
It therefore makes the application of the GFBM to the systems of interest a very difficult task,  as a very large number of breakup channels has to be taken into account for large and highly excited systems, such as those considered in the multifragment emission \cite{Bondorf1995,BettyPhysRep2005}.
Since in the framework of the GFBM the SMM should be repeatedly applied to calculate the decay of the hot fragments, the application of the model to large systems and high excitation energies becomes almost prohibitively time consuming.

In this work, we present an implementation of the GFBM based on the exact recursion formulae developed by Chase and Mekjian \cite{ChaseMekjian1995} and further developed in subsequent works \cite{SubalMekjian,Subal1999,PrattDasGupta2000,BettyPhysRep2005}  to calculate the partition function.
Since this scheme allows the calculation of the statistical weights to be performed with high efficiency, it turns out to be well suited to speeding up the very computational demanding calculations needed in the GFBM.

We thus start, in Sec.\ \ref{sect:primary}, by reviewing the SMM, on which the model is based, and also discuss the recursion relations just mentioned. 
A comparison of the results obtained with the Monte-Carlo version of the SMM and that based on these recursion formulae, which we name SMM with Discrete Energy (SMM-DE), is also made.
Then, in Sec.\ \ref{sect:deexcit} we work out the implementation of the GFBM and discuss some results.
Concluding remarks are drawn in Sec.\ \ref{sect:conclusions}.

\end{section}

\begin{section}{The primary fragments}
\label{sect:primary}
In the framework of the SMM \cite{smm1,smm2,smm4}, an excited source, with $A_0$ nucleons, $Z_0$ protons, and excitation energy $E^*$, undergoes a prompt breakup. 
Hot primary fragments are thus produced, except for those which have no internal degrees of freedom and are therefore cold, {\it i.e.} nuclei whose mass number $A\le 4$, except for $\alpha$ particles.
Partitions  consistent with mass, charge, and energy conservation are generated and the corresponding statistical weights are calculated.
In Sec.\ \ref{sect:primSMM}, we briefly review the main features of the SMM.
The implementation based on the discretization of the energy, using the recurrence relations developed by Pratt and Das Gupta \cite{PrattDasGupta2000} is discussed in Sec.\ \ref{sect:SMMDE}.
A comparison between the two implementations is made in Sec.\ \ref{sect:compSMM}.

\begin{subsection}{The SMM model}
\label{sect:primSMM}
In each fragmentation mode $f$, the multiplicity $n_{A,Z}$ of a species, respectively possessing mass and atomic numbers $A$ and $Z$, must fulfill the constraints:

\begin{equation}
A_0=\sum_{\{A,Z\}}A\,n_{A,Z} \;\;\;\;\;{\rm and}\;\;\;\; Z_0=\sum_{\{A,Z\}}Z\,n_{A,Z}\;.
\label{eq:cmCons}
\end{equation}

\noindent
The statistical weight $\Omega_f$ associated with the fragmentation mode $f=\{n_{A,Z}\}$ is given by the number of micro-states associated with it:

\begin{equation}
\Omega_f(E)=\exp(S_f)\;,
\label{eq:ws}
\end{equation}

\noindent
where $E=-B_{A_0,Z_0}+E^*$ denotes the total energy of the system, $B_{A_0,Z_0}$ represents the binding energy of the source, and  $S_f$ corresponds to the total entropy of the fragmentation mode:

\begin{equation}
S_f=\sum_{\{A,Z\}}n_{A,Z}S_{A,Z}\;.
\label{eq:stotal}
\end{equation}

\noindent
The entropy  $S_{A,Z}$ is obtained through the standard thermodynamical relation:

\begin{equation}
S=-\frac{dF}{dT}\;.
\label{eq:sdfdt}
\end{equation}

\noindent
In the above expression, $T$ denotes the breakup temperature and $F_{A,Z}$ is the Helmholtz free energy associated with the species, which is related to its energy $E_{A,Z}$ through

\begin{equation}
E=F+TS\;.
\label{eq:fee}
\end{equation}

\noindent
The breakup temperature of the fragmentation mode $f$ is obtained through the energy conservation constraint:

\begin{equation}
E^*-B_{A_0,Z_0}=C_c\frac{Z_0^2}{A_0^{1/3}}\frac{1}{(1+\chi)^{1/3}}+\sum_{\{A,Z\}}n_{A,Z}E_{A,Z}(T_f)\;,
\label{eq:econst}
\end{equation}

\noindent
where the subindex $f$ was used in $T_f$ to emphasize the fact that the breakup temperature varies from one partition to another \cite{thermometry2000}, although we drop this subindex from now on to simplify the notation.
In the above equation, the first term on the right hand side  corresponds to the Coulomb energy of a uniform sphere of volume $V=(1+\chi)V_0$, $\chi > 0$, where $V_0$ denotes the ground state volume of the source and $C_c$ is a parameter (see below).
In this work, we use $\chi=2$ in all calculations.
The fragment energy $E_{A,Z}(T)$ reads:

\begin{equation}
E_{A,Z}(T)=-B_{A,Z}+\epsilon^*_{A,Z}-C_c\frac{Z^2}{A^{1/3}}\frac{1}{(1+\chi)^{1/3}}
+E_{A,Z}^{\rm trans}\;,
\label{eq:eaz}
\end{equation}

\noindent
where $B_{A,Z}$ stands for the fragment's binding energy, $\epsilon^*_{A,Z}$ denotes the internal excitation energy of the fragment, and the contribution $E_{A,Z}^{\rm trans}$ to the total kinetic energy $E^{\rm trans}$ reads

\begin{equation}
E^{\rm trans}=\sum_{\{A,Z\}}n_{A,Z} E_{A,Z}^{\rm trans}=\frac{3}{2}(M-1)T
\label{eq:etrans}
\end{equation}

\noindent
where $M=\sum_{\{A,Z\}}n_{A,Z}$ is the total multiplicity of the fragmentation mode $f$.
The factor $M-1$, rather than $M$, takes into account the fact that the center of mass is at rest.
Together with the Coulomb terms in the fragments' binding energies and that of the homogeneous sphere in Eq.\ (\ref{eq:econst}), the Coulomb contribution on the right hand side of Eq.\ (\ref{eq:eaz}) adds up to account for the Coulomb energy of the fragmented system in the Wigner-Seitz approximation \cite{smm1,WignerSeitz}.

In Ref.\ \cite{ISMMlong}, empirical values were used for $B_{A,Z}$ and an extrapolation scheme was developed to the mass region where experimental information is not available. 
For simplicity, in this work, except for $A\le 4$, in which case empirical values are used, we adopt the mass formula developed in Ref.\ \cite{ISMMmass}:

\begin{equation}
B_{A,Z}=C_vA-C_sA^{2/3}-C_c\frac{Z^2}{A^{1/3}}+C_d\frac{Z^2}{A}+\delta_{A,Z}A^{-1/2}\;,
\label{eq:be}
\end{equation}

\noindent
where

\begin{equation}
C_i=a_i\left[1-k\left(\frac{A-2Z}{A}\right)^2\right]\;.
\end{equation}

\noindent
and $i=v,s$ respectively denotes the volume and surface terms.
The last term in Eq.\ (\ref{eq:be}) is the usual pairing contribution:

\begin{equation}
\delta_{A,Z}=\frac{1}{2}\left[(-1)^{A-Z}+(-1)^Z\right]C_p\;.
\label{eq:delta}
\end{equation}

\noindent
We refer the reader to Ref.\ \cite{ISMMmass} for numerical values of the parameters.

In the standard version of SMM \cite{smm1}, the contribution to the entropy and the fragment's energy due to the internal degrees of freedom is obtained from the internal Helmholtz free energy:

\begin{equation}
F_{A,Z}^*(T)=-\frac{T^2}{\epsilon_0}A+\beta_0A^{2/3}\left[\left(\frac{T_c^2-T^2}{T_c^2+T^2}\right)^{5/4}-1\right]\;,
\label{eq:feex}
\end{equation}

\noindent
where $T_c=18.0$ MeV, $\beta_0=18.0$ MeV, and $\epsilon_0=16.0$ MeV.
In Ref.\ \cite{ISMMlong}, effects associated with discrete excited states have been incorporated into $F_{A,Z}^*(T)$.
For simplicity, in the present work, we only use the above expression.

Finally, the total contribution to the entropy associated with the translational motion reads:

\begin{eqnarray}
& & F_{\rm trans}=-T(M-1)\log(V_f/\lambda_T^3)+T\log(A_0^{3/2})\nonumber\\
&-&T\sum_{\{A,Z\}}n_{A,Z}\left[\log(g_{A,Z}A^{3/2})-\frac{1}{n_{A,Z}}\log(n_{A,Z}!)\right]\;.
\label{eq:ftrans}
\end{eqnarray}

\noindent
In the above expression, $V_f=\chi V_0$ denotes the free volume and the factor $M-1$, as well as the term $T\log(A_0^{3/2})$, arise from the constraint that the center of mass be at rest \cite{isoSMMTF,fbk2012}.
The thermal wave length reads $\lambda_T=\sqrt{2\pi\hbar^2/m_nT}$, where $m_n$ is the nucleon mass.
Empirical values of the spin degeneracy factors $g_{A,Z}$ are used for $A\le 4$.
In the case of heavier nuclei, we set $g_{A,Z}=1$ as (to some extent) this  is taken into account by $F^*_{A,Z}$.

Fragmentation modes are generated by carrying out the following steps:

\begin{description}
\item{\bf i-} The multiplicities $\{n_{A,Z}\}$ are sampled under the constraints imposed by Eq.\ (\ref{eq:cmCons}), as described in Ref.\ \cite{smm4};
\item{\bf ii-} Equation\ (\ref{eq:econst}) is solved in order to determine the breakup temperature $T$;
\item{\bf iii-} The total entropy is calculated through Eqs.\ (\ref{eq:stotal}) and (\ref{eq:sdfdt}), after having computed the Helmholtz free energies from Eqs.\ (\ref{eq:feex}) and (\ref{eq:ftrans}).
\end{description}

\noindent
The average value of an observable $O$ is then calculated through:

\begin{equation}
\overline{O}=\frac{\sum_f \Omega_f(E) O_f}{\sum_f \Omega_f(E)}\;.
\label{eq:aveO}
\end{equation}

\noindent
Since hundreds of million partitions must be generated in order to achieve a reasonable sampling of the available phase space, this implementation, although feasible, is time consuming.
It has been quite successful in describing several features of the multifragmentation process \cite{Bondorf1995}.

\end{subsection}

\begin{subsection}{The SMM-DE model}
\label{sect:SMMDE}
A much more efficient scheme has been proposed by Chase and Mekjian \cite{ChaseMekjian1995}, who developed an exact method based on recursion formulae.
This allows one to easily compute the number of states $\Omega_A$ associated with the breakup of a nucleus of mass number $A$  in the canonical ensemble through \cite{SubalMekjian}:

\begin{equation}
\Omega_A=\sum_{\{\sum_k n_k a_k=A\}}\prod_k\frac{\omega_k^{n_k}}{n_k!}=\sum_{a=1}^A\frac{a}{A}\omega_a\Omega_{A-a}\;,
\label{eq:omegaA}
\end{equation}

\noindent
where $\omega_k$ denotes the number of states of a nucleus of mass number $a$.

This result has  been later generalized to distinguish protons from neutrons, leading to a similar expression \cite{Subal1999}: 

\begin{eqnarray}
\Omega_{A,Z}&=&\sum_{\{\sum_{\alpha}n_{a_\alpha,z_\alpha}\zeta_\alpha=\Lambda\}}\prod_\alpha\frac{(\omega_{a_\alpha,z_\alpha})^{n_{a_\alpha,z_\alpha}}}{n_{a_\alpha,z_\alpha}!}
\nonumber\\
&=&\sum_{\alpha}\frac{\zeta_\alpha}{\Lambda}\omega_{a_\alpha,z_\alpha}\Omega_{A-a_\alpha,Z_-z_\alpha}\;,
\end{eqnarray}

\noindent
where $\Lambda$ denotes $A$ or $Z$, and $\zeta_\alpha$ conveniently represents either $a_\alpha$ or $z_\alpha$.
Although somewhat more involved than the previous expression, it still allows one to  calculate the statistical weight associated with the breakup of a nucleus $(A,Z)$ very efficiently, as well as other average quantities, such as the average multiplicities \cite{Subal1999}:

\begin{eqnarray}
\overline{n}_{a,z}&=&\frac{1}{\Omega_{A,Z}}\sum_{\{\sum_{\alpha}n_{a_\alpha,z_\alpha}\zeta_\alpha=\Lambda\}}n_{a,z}\prod_\alpha\frac{(\omega_{a_\alpha,z_\alpha})^{n_{a_\alpha,z_\alpha}}}{n_{a_\alpha,z_\alpha}!}\nonumber\\
&=&\frac{\omega_{a,z}}{\Omega_{A,Z}}\Omega_{A-a,Z-z}\;.
\label{eq:aveNaz}
\end{eqnarray}

\noindent
This scheme has been successfully applied to the description of the breakup of excited nuclear systems during the last decade \cite{BettyPhysRep2005}.

The extension to the microcanonical ensemble has been developed in Ref.\ \cite{PrattDasGupta2000}.
More specifically, Eq.\ (\ref{eq:econst}) can be rewritten as:

\begin{equation}
Q\Delta_Q\equiv E^*-B^c_{A_0,Z_0}=\sum_{\alpha,q_\alpha} q_\alpha n_{\alpha,q_\alpha} \;,
\label{eq:qconst}
\end{equation}

\noindent
where $q_{\alpha}\Delta_Q$ denotes the fragment's energy, together with the corresponding Wigner-Seitz contribution to the Coulomb energy:

\begin{equation}
q_{A,Z}\Delta_Q=-B^c_{A,Z}+\epsilon^*_{A,Z}+E_{A,Z}^{\rm trans}
\label{eq:qaz}
\end{equation}

\noindent
and

\begin{equation}
B^c_{A,Z}\equiv B_{A,Z}+C_c\frac{Z^2}{ A^{1/3}}\frac{1}{(1+\chi)^{1/3}}\;.
\label{eq:bcaz}
\end{equation}

\noindent
In contrast to the continuous quantities $E$ and $E_{A,Z}$, $Q$ and $q_{A,Z}$ are discrete.
The granularity of the discretization is conveniently regulated by the energy bin $\Delta_Q$.
In this way, $Q$ may be treated as a conserved quantity, similarly to the mass and atomic numbers, so that the recursion relation now reads \cite{PrattDasGupta2000}:

\begin{equation}
\Omega_{A,Z,Q}=\sum_{\alpha,q_\alpha}\frac{a_\alpha}{A}\omega_{a_\alpha,z_\alpha,q_{\alpha}}\Omega_{A-a_\alpha,Z-z_\alpha,Q-q_{\alpha}}\;,
\label{eq:oazq}
\end{equation}

\noindent
and the average multiplicity is given by

\begin{equation}
\overline{n}_{a,z,q}=\frac{\omega_{a,z,q}}{\Omega_{A_0,Z_0,Q}}\Omega_{A_0-a,Z_0-z,Q-q}\;.
\label{eq:nazq}
\end{equation}

\noindent
Other average quantities, such as the breakup temperature and the total entropy, can also be readily obtained:

\begin{equation}
\frac{1}{T}=\frac{\partial\ln(\Omega_{A_0,Z_0,Q})}{\partial (Q\Delta_Q)}\approx \frac{\ln(\Omega_{A_0,Z_0,Q})-\ln(\Omega_{A_0,Z_0,Q-1})}{\Delta_Q}
\label{eq:disc}
\end{equation}

\noindent
and

\begin{equation}
S=\ln(\Omega_{A_0,Z_0,Q})\;.
\label{eq:sdisc}
\end{equation}

The statistical weight $\{\Omega_{A,Z,q}\}$ is determined once $\omega_{A,Z,q}$ has been specified.
The latter is obtained by folding the number of states associated with the kinetic motion with that corresponding to  the internal degrees of freedom:

\begin{equation}
w_{A,Z,q}=\gamma_A\int_0^{\epsilon_{A,Z,q}}\,dK\;\sqrt{K}\rho^*_{A,Z}(\epsilon_{A,Z,q}-K)\;,
\label{eq:wazq}
\end{equation}

\noindent
where

\begin{equation}
\gamma_A=\Delta_Q \frac{V_f (2m_n A)^{3/2}}{4\pi^2\hbar^3}\;,
\label{eq:gamma}
\end{equation}

\noindent
$\epsilon_{A,Z,q}\equiv {\,q\Delta_Q+B^c_{A,Z}}$, and $\rho^*_{A,Z}(\epsilon^*)$ is the density of the internal states of the nucleus $(A,Z)$ with excitation energy $\epsilon^*$.
It thus becomes clear that the fundamental physical ingredient is $\rho^*_{A,Z}(\epsilon^*)$ as it plays a major role in the determination of the statistical weight.

Since $\Omega_{A,Z,q}$ depends on three variables and the calculation of $\omega_{A,Z,q}$ usually must be evaluated numerically, the computation of $\Omega_{A_0,Z_0,Q}$ may be very time consuming, for big sources, large excitation energies and small values of $\Delta_Q$ (large $Q$).
Nevertheless, $\omega_{A,Z,q}$ and $\Omega_{A,Z,q}$ need to be calculated only once and may be stored, in order to considerably speed up future  calculations, even for different sources as these quantities do not depend on the source's properties.
This is the strategy we adopt.

In Ref.\ \cite{ISMMlong}, it has been shown that the standard SMM internal free energies is fairly well approximated in a wide range of temperatures if one adopts the following density of states:

\begin{equation}
\rho^*_{A,Z}(\epsilon^*)=\rho_{\rm SMM}(\epsilon^*)=\rho_{\rm FG}(\epsilon^*)e^{-b_{\rm SMM}(a_{\rm SMM}\epsilon^*)^{3/2}}
\label{eq:rhoSMM}
\end{equation}

\noindent
with

\begin{equation}
\rho_{\rm FG}(\epsilon^*)=\frac{a^{1/4}_{\rm SMM}}{\sqrt{4\pi}{\epsilon^*}^{3/4}}\exp(2\sqrt{a_{\rm SMM}\epsilon^*})
\label{eq:rhofg}
\end{equation}

\noindent
and

\begin{equation}
a_{\rm SMM}=\frac{A}{\epsilon_0}+\frac{5}{2}\beta_0\frac{A^{2/3}}{T_c^2}\;.
\label{eq:asmm}
\end{equation}

\noindent
The parameter $b_{\rm SMM}=0.07 A^{-\tau}$, $\tau=1.82(1+A/4500)$, for $A>4$.
In the case of the alpha particles, we set $\beta_0=0$ and $b_{\rm SMM}=0.000848416$.
For the other light nuclei whose $A<5$, which have no internal degrees of freedom, we use $\rho^*_{A,Z}(\epsilon^*)=g_{A,Z}\delta(\epsilon^*)$.

As mentioned above, we name this implementation of the model SMM-DE.

\end{subsection}

\begin{subsection}{Comparison with the SMM}
\label{sect:compSMM}
In order to investigate the extent to which the above implementations of the SMM agree with each other, we show in Fig.\ \ref{fig:cc} the caloric curve predicted by both versions of SMM for the breakup of the $^{40}$Ca nucleus.
As expected, at low temperatures, the caloric curve is very close to that of a Fermi-gas but this behavior quickly changes for $E^*/A\gtrsim 2$ MeV, as the fragment multiplicity $M$ starts to deviate from one (see top panel of Fig.\ \ref{fig:multip}).
From this point on the temperature rises slower than that of a Fermi gas as the excitation energy increases, due to the larger heat capacity of the system which progressively produces more fragments with internal degrees of freedom, besides very light ones.

\begin{figure}[tbh]
\includegraphics[width=8.5cm,angle=0]{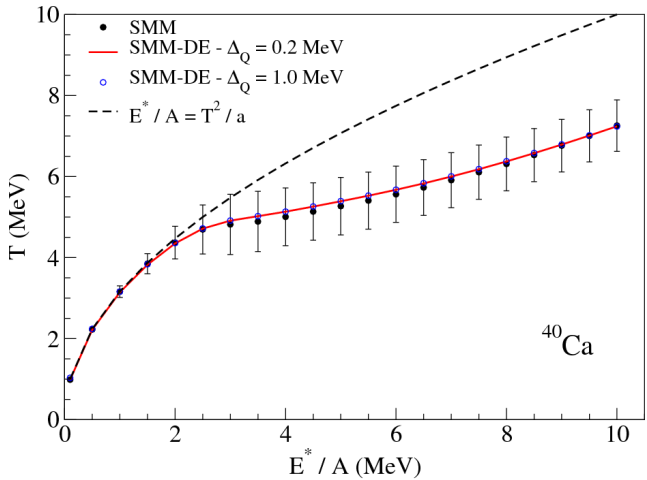}
\caption{\label{fig:cc} (Color online) Caloric curve predicted by the SMM and the SMM-DE for the $^{40}$Ca. The dashed line represents the Fermi-gas with $a=10$ MeV. For details see the text.}
\end{figure}

Comparison between the results obtained with the SMM and the SMM-DE shows that both versions predict very similar caloric curves.
To some extent, slight deviations should be expected since there are small differences between the two implementations.
Firstly, in the SMM the constraint on the center of mass motion is taken into account by Eqs.\ (\ref{eq:etrans}) and (\ref{eq:ftrans}), whereas it is ignored in the SMM-DE.
Therefore, the latter has more states than the former, leading to different statistical weights and, as a consequence, different averages.
Secondly, as pointed out in Ref.\ \cite{PrattDasGupta2000}, for a given fragmentation mode, fluctuations around the mean excitation energy  of each fragment are not allowed in the SMM, whereas the summation over $q$ in Eq.\ (\ref{eq:oazq}) considers all possible ways of sharing the energy amongst the fragments.
The results nonetheless reveal that the practical consequences of these two points are small.

The comparison between the results obtained with $\Delta_Q=0.2$ MeV and $\Delta_Q=1.0$ MeV, also displayed in Fig.\ \ref{fig:cc}, shows that this parameter should have little influence on the results.
This is of particular relevance since the numerical effort increases very fast as $\Delta_Q$ decreases, which makes the application of the model for big systems at large excitation energies much more time consuming.
We use $\Delta_Q=0.2$ MeV in the remainder of this work, but these results suggest that somewhat larger values would be just as good.

\begin{figure}[tbh]
\includegraphics[width=8.5cm,angle=0]{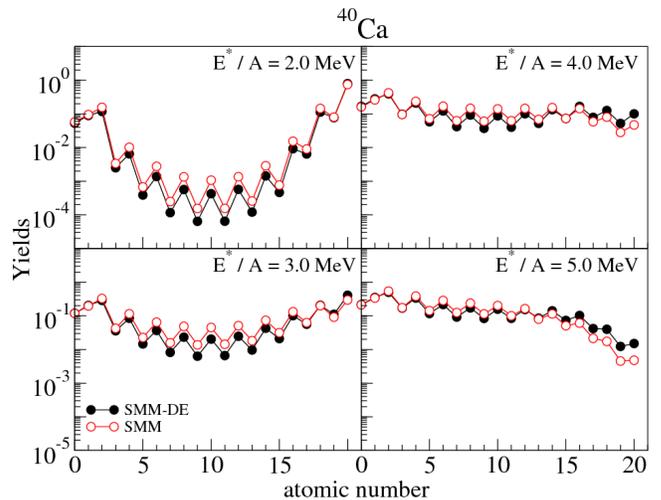}
\caption{\label{fig:yzprim} (Color online) Charge distribution of the hot primary fragments from the breakup of the $^{40}$Ca, obtained with the different versions of SMM at selected excitation energies. For details see the text.}
\end{figure}

More detailed information on the similarities of the two versions of the SMM may be obtained by examining the charge distribution of the primary fragments produced at different excitation energies, as  shown in Fig.\ \ref{fig:yzprim}.
Once also sees in this case that both SMM calculations lead to very similar predictions, although small deviations are observed.
They are more pronounced in the $Z$ region close to the source size since its contribution is overestimated in the SMM-DE.
However, the differences in lower $Z$ regions are smaller and should not impact on the conclusions drawn from either implementation, within the usual uncertainties of these calculations.
We therefore believe that either implementations of the SMM can be safely adopted.

\end{subsection}

\end{section}

\begin{section}{The deexcitation of the primary fragments}
\label{sect:deexcit}
In Ref.\ \cite{fbk2012}, it has been demonstrated that the statistical description of the multifragment emission made by the SMM is equivalent to a generalized version of the Fermi Breakup, in which the excited states of the fragments are  included.
We here  pursue this idea  and apply it to the description of the deexcitation of the hot primary fragments, referring the reader to that work for a detailed discussion.

The starting point is Eq.\ (\ref{eq:nazq}) which provides the average multiplicity $\overline{n}_{a,z,q}$ of a fragment $(a,z)$ with total energy $q\Delta_Q$, produced in the breakup of a source $(A_0,Z_0)$ with total energy $Q\Delta_Q$.
The average excitation energy $\overline{\epsilon}^*$ of the fragment is calculated through:

\begin{equation}
\overline{\epsilon}^*=\frac{\gamma_a}{\omega_{a,z,q}}\int_0^{\epsilon_{a,z,q}}\,dK\;(\epsilon_{a,z,q}-K)\sqrt{K}\rho^*_{a,z}(\epsilon_{a,z,q}-K)\;.
\label{eq:eex}
\end{equation}

In Fig.\ \ref{fig:eex} we denote by $n(\overline{\epsilon}^*)$  the average multiplicity of C isotopes with energy $q\Delta_Q$, {\it i.e.} $\overline{n}_{A,6,q}$, and plot this quantity, scaled by the corresponding maximum value, as a function of the average excitation energy $\overline{\epsilon}^*$.
The results were obtained from the breakup of $^{40}$Ca at $E^*/A=3$, 5 and 7 MeV.
One notes that $\overline{\epsilon}^*$ is very broad and that the width of the distribution becomes larger as $E^*/A$ increases.
The average value of $\overline{\epsilon}^*$ also  shifts to higher values as the excitation energy of the source $E^*/A$ becomes larger.
It is clear from these results that the internal excitation of the fragments cannot be neglected and also that the width of the distribution is by no means negligible, even at low excitation energies.
The inset of the figure illustrates the isospin dependence of the distribution by displaying $n(\overline{\epsilon}^*)$ for different C isotopes also produced in the breakup of $^{40}$Ca at
$E^*/A=5.0$ MeV.
It shows that the proton rich isotope is cooler than the neutron rich one which, in its turn, is slightly hotter than $^{12}$C.
Similar conclusions hold for other species.

\begin{figure}[tbh]
\includegraphics[width=8.5cm,angle=0]{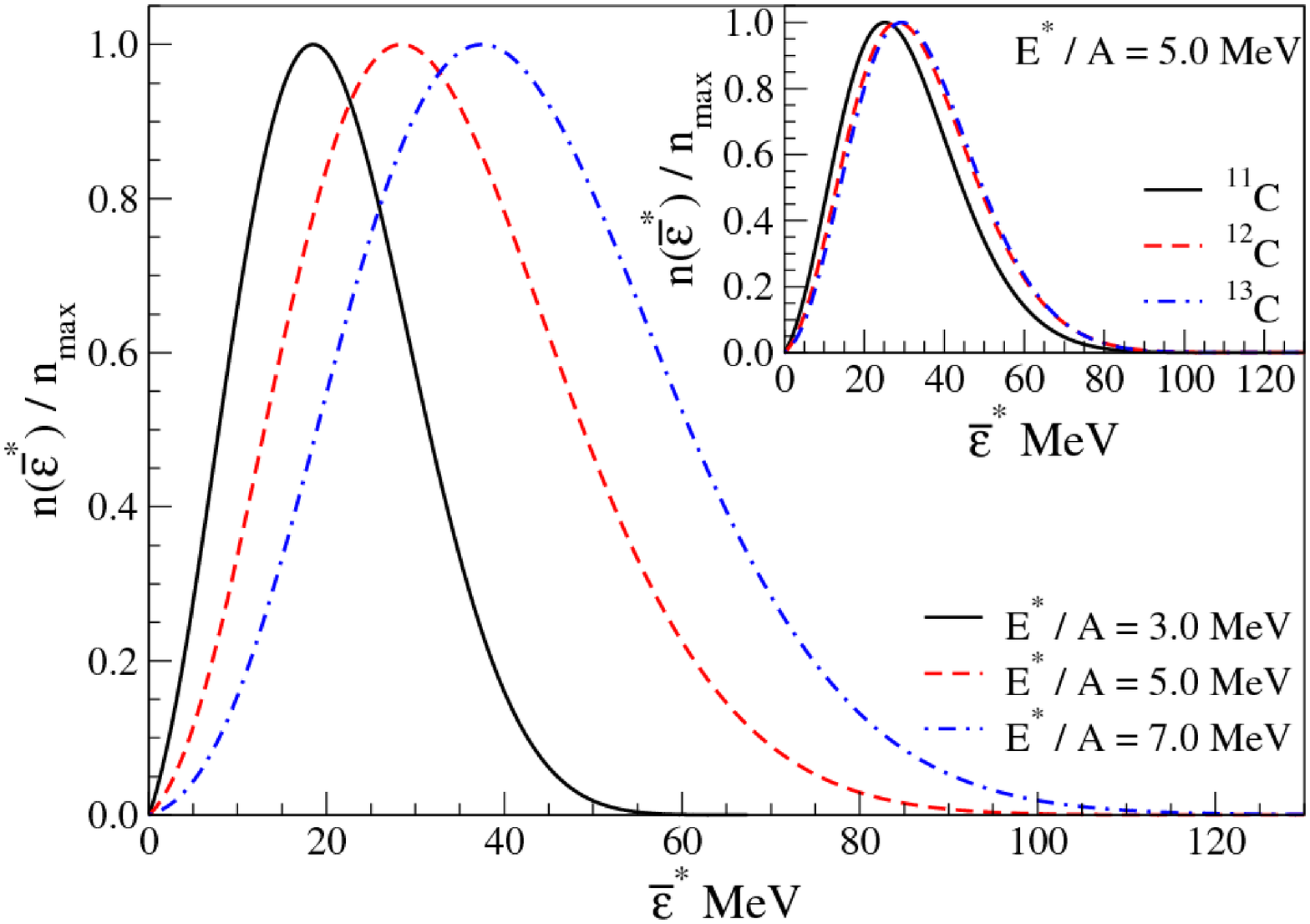}
\caption{\label{fig:eex} (Color online) Multiplicity of hot primary $^{12}$C  as a function of its average excitation energy $\overline{\epsilon}^*$.
The fragments are produced in the breakup of $^{40}$Ca at different excitation energies, as indicated in the legend. The inset shows the same plot for different C isotopes. In this case, the source energy is $E^*/A = 5.0$ MeV. The multiplicities have been scaled by the inverse of the largest value in each case. For details see the text.}
\end{figure}

Since they are hot, these  primary  fragments are themselves considered as sources and are then allowed to deexcite through the procedure  starting at Eq.\ (\ref{eq:qconst}).
Given that, in actual experiments, almost all fragments have already decayed by the time they reach to the detectors, we apply this procedure successively until the fragment is left in its ground state, if it is stable and does not spontaneously decay by the emission of lighter fragments.
In this case, this is treated using the same formalism presented above, with $E^*=0$.

Thus, once the primary distribution $\{n_{a,z,q}\}$ has been generated, the deexcitation of the fragments follows the steps below:

\begin{description}
\item{\bf i-} The average excitation energy $\overline{\epsilon}^*$ of a fragment $(A,Z)$ with energy $q_0\Delta_Q$ is calculated from Eq.\ (\ref{eq:eex}) and the decay described in Sec.\ \ref{sect:SMMDE} is applied to it.
The corresponding contribution to the yields of $\{a,z\}$, based on Eq.\ (\ref{eq:nazq}), {\it i.e.}

\begin{equation}
\Delta \overline{n}_{a,z,q}^{(1)}=\overline{n}'_{a,z,q}\times\overline{n}_{A,Z,q_0}
\label{eq:deltanazq1}
\end{equation}

\noindent
where
\begin{equation}
\overline{n}'_{a,z,q}=\frac{\omega_{a,z,q}}{\Omega_{A,Z,q_0}}\Omega_{A-a,Z-z,q_0-q}\;,
\label{eq:nazqp}
\end{equation}

\noindent
is added to $\overline{n}_{a,z,q}$, $a < A$.

\item{\bf ii-} Since a fragment $(A,Z)$ with energy $q_0\Delta_Q$ will also be produced at this stage, with multiplicity  $\overline{n}'_{A,Z,q_0}\times\overline{n}_{A,Z,q_0}$, it will again decay and contribute to the yields of lighter fragments $(a,z)$ in the second step with

 \begin{equation}
\Delta \overline{n}_{a,z,q}^{(2)}=\overline{n}'_{a,z,q}\times(\overline{n}'_{A,Z,q_0}\times\overline{n}_{A,Z,q_0})
\label{eq:deltanazq2}
\end{equation}

whereas there will still be a contribution to the yields of $(A,Z)$ equal to $(\overline{n}'_{A,Z,q_0})^2\times\overline{n}_{A,Z,q_0}$.
Thus, the $n$-{\it th} step of the decay contributes with

 \begin{equation}
\Delta \overline{n}_{a,z,q}^{(n)}=\overline{n}'_{a,z,q}\times([\overline{n}'_{A,Z,q_0}]^{n-1}\times\overline{n}_{A,Z,q_0})\;.
\label{eq:deltanazqn}
\end{equation}

Since these terms add up at each step, after repeating this procedure until the contribution to $(A,Z)$ tends to zero, {\it i.e.} $n\rightarrow\infty$, one obtains:

\begin{equation}
\overline{n}_{a,z,q}\rightarrow\overline{n}_{a,z,q}+\frac{\overline{n}'_{a,z,q}}{1-\overline{n}'_{A,Z,q_0}}\times\overline{n}_{A,Z,q_0}\;,\;\;\; a < A\;.
\label{eq:yrenrom}
\end{equation} 

\item{\bf iii-} After carrying out steps (i) and (ii) for all the isobars $A$, one decrements $A$ by one unity and goes back to step (i), until all the excited fragments have decayed.

\end{description}

\noindent
In order to speed up the calculations, the distribution $n(\overline{\epsilon}^*)$ of the average excitation energies $\overline{\epsilon}^*$  of the decaying fragment (see Fig.\ \ref{fig:eex}) is binned in bins of size $\Delta\overline{\epsilon}^*$, for $\overline{\epsilon}^*>1.0$ MeV.
Very low excitation energies, {\it i.e.} $0\le\overline{\epsilon}^*\le 1.0$ MeV, are always grouped in a bin of size $1.0$ MeV, regardless of the value of $\Delta\overline{\epsilon}^*$ employed in the calculation.

In Fig.\ \ref{fig:multip} we show the total multiplicity and the number of intermediate mass fragment  ($N_{\rm IMF}$) as a function of the total excitation energy  $E^*/A$ of the $^{40}$Ca source, for both primary and final yields, using $\Delta\overline{\epsilon}^*=1$ MeV.
One sees that the total primary multiplicity $M$, shown in the upper panel, rises steadily as $E^*$ increases, for $E^*/A\gtrsim 2$ MeV.
Up to this point,  $M$ is close to unity, which means that the excited source does not decay in the primary stage.
On the other hand, when the deexcitation scheme just described is applied to the primary fragments, one also sees in the top panel of Fig.\ \ref{fig:multip} that $M$ increases continuously.
It should be noted that most of the fragments are produced in the deexcitation stage.
This suggests that the relevance of this stage of the reaction is, at least, as important as the prompt breakup for this copious secondary  particle emission can hide the underlying physics governing the primary stage.

\begin{figure}[tbh]
\includegraphics[width=8.5cm,angle=0]{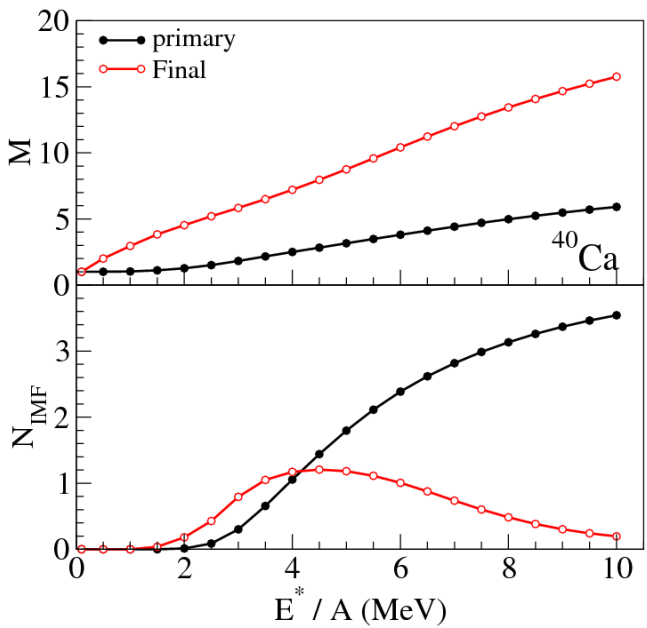}
\caption{\label{fig:multip} (Color online) Top panel: Total fragment multiplicity before and after secondary decay as a function of the total excitation energy of the source. Bottom Panel: IMF multiplicity ($3\le Z \le 15$) from the hot primary fragments and after the deexcitation process. For details see the text.}
\end{figure}

The lower panel of Fig.\ \ref{fig:multip} displays the multiplicity of Intermediate Mass Fragments (IMF), $N_{\rm IMF}$, as a function of the excitation energy
obtained with both the primary and the final yields.
It is built by adding up the multiplicities $\overline{n}_{a,z}$ whose $3 \le Z\le 15$.
The results show that the primary $N_{\rm IMF}$ is zero for $E^*/A\lesssim 2$ MeV since, up to this point, only the heavy remnant is present, but it quickly departs from zero at that point, rising steadily from there on.
On the other hand, the final $N_{\rm IMF}$ first rises and then falls off since, although fragments with $Z\ge 3$ are also produced in the secondary stage, many IMF's are  destroyed as they tend to emit more and more very light fragments ($Z\le 2$) as the excitation energy increases.
These results are in qualitative agreement with those presented in Refs.\ \cite{Bondorf1995,ISMMlong}, obtained with different treatments.

\begin{figure}[tbh]
\includegraphics[width=8.5cm,angle=0]{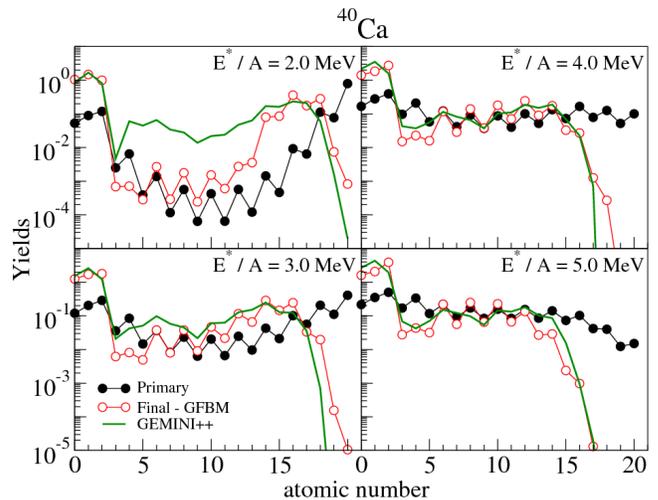}
\caption{\label{fig:zDist} (Color online) Charge distribution of fragments produced in the breakup of $^{40}$Ca at selected excitation energies.
The primary distributions are obtained with the SMM-DE, whereas the final yields are calculated with the GFBM presented in this work.
The predictions of the GEMINI++ code for the decay of a compound nucleus are also displayed. For details see the text.}
\end{figure}

The comparison between the primary and final yields respectively obtained with the SMM-DE and the GFBM is shown in Fig.\ \ref{fig:zDist}.
One observes that the qualitative shape of the charge distribution does not change appreciably after the deexcitation of the primary hot fragments, but the suppression of heavy residues becomes progressively more important as the excitation energy increases, while the yields of the light fragments are enhanced.
These results show that, in the excitation energy domain studied in this work, there are important quantitative differences between the primary and final charge distributions.

In order to investigate the extent to which our model agrees with others traditionally used in this energy domain, we also display in Fig.\ \ref{fig:zDist} the results obtained with the GEMINI++ code \cite{GEMINI1988,GEMINIpp2010_1,GEMINIpp2010_2}.
We considered the reaction $^{20}{\rm Ne}+{^{20}{\rm Ne}}$, at the appropriate bombarding energy, leading to a compound nucleus equal to $^{40}$Ca with the suitable excitation energy.
One observes a very good agreement with the predictions made by the GEMINI++ code and the GFBM, which systematically improves as the excitation energy of the source increases.
This is probably due to the different assumptions made by the two models, {\it i.e.} binary emission versus prompt breakup, which seem to affect the charge distribution only at very low excitation energies.

Finally, we examine the sensitivity of the model results to the binning used to speed up the calculations in the secondary decay stage.
The charge distributions for the breakup of $^{40}$Ca at $E^*/A=5.0$ MeV is displayed in Fig.\ \ref{fig:zDistDeltaE} for $\Delta\overline{\epsilon}^*=1.0$, 5.0, and 10.0 MeV.
It is clear that the charge distribution is weakly affected by this parameter so that values within this range may be safely used, as the variations are within the model's precision.

\begin{figure}[tbh]
\includegraphics[width=8.5cm,angle=0]{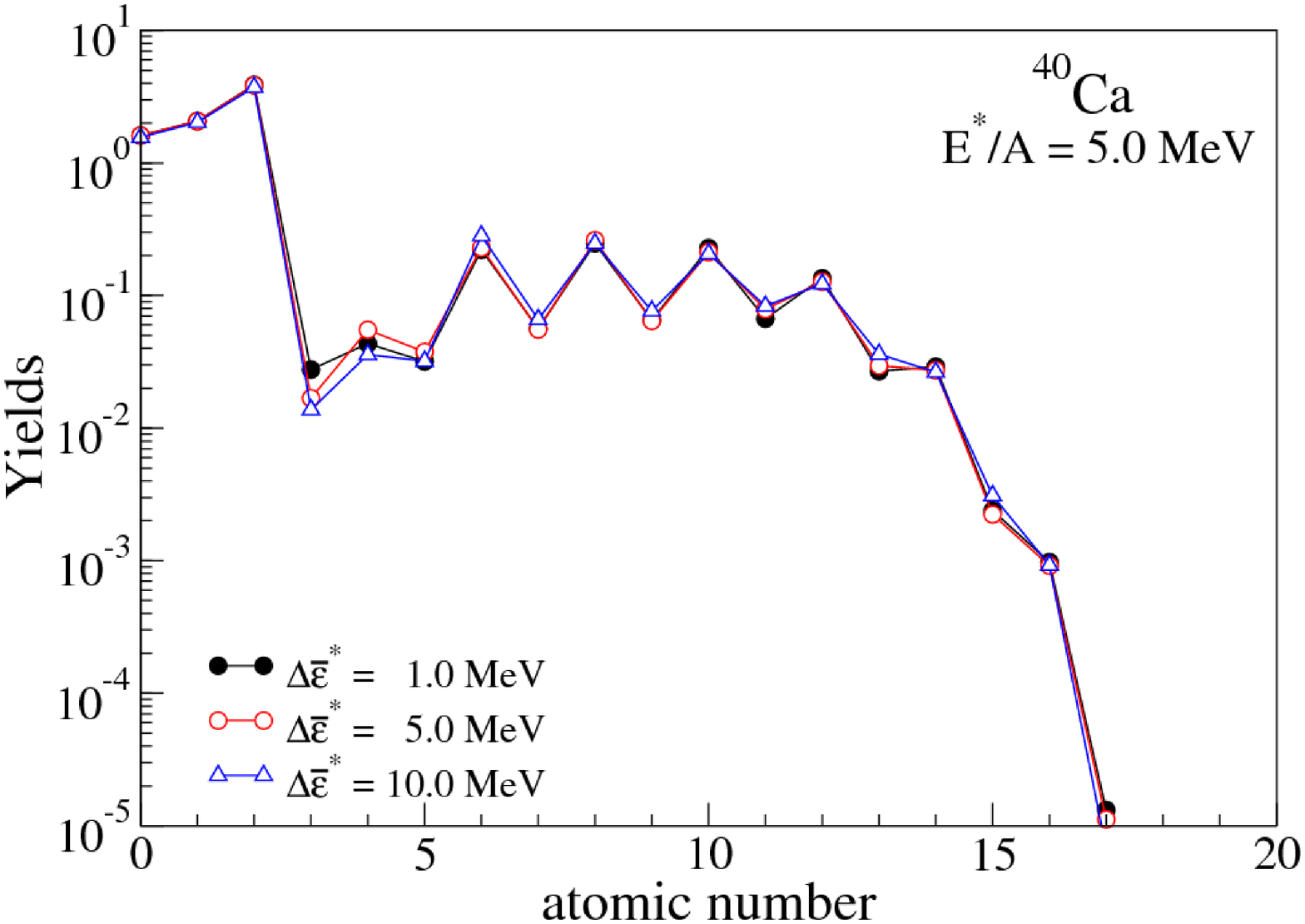}
\caption{\label{fig:zDistDeltaE} (Color online) Charge distribution of fragments produced in the breakup of $^{40}$Ca at $E^*/A=5.0$ MeV.
All the parameters of the calculation are kept fixed, except for $\Delta\overline{\epsilon}^*$.
For details see the text.}
\end{figure}

\end{section}

\begin{section}{Concluding remarks}
\label{sect:conclusions}
We have presented an implementation of the Generalized Fermi Breakup model, introduced in Ref.\ \cite{fbk2012}, to treat the deexcitation of the primary hot fragments produced in the breakup of a nuclear source.
The approach is based on the SMM \cite{smm1,smm2,smm4,ISMMmass}, which describes the primary breakup stage.
It is then successively applied to the primary fragments until they have decayed to the ground state.
Since the application of the SMM to all the primary fragments, repeatedly until they are no longer excited, would be extremely time consuming, we have developed an implementation of the SMM based on the recursion formulae presented in Ref.\ \cite{PrattDasGupta2000}.
Those formulae allow the statistical weights to be very efficiently computed so that they make the application of our model to systems of interest feasible.
We found that the traditional Monte Carlo implementation of SMM and that developed in the present work lead to very similar results, so that either one may be used according to the need.
Furthermore, the similarity of the final yields obtained with our GFBM with those predicted by the GEMINI++  code strongly suggests that our treatment is, at least, as good as the more traditional ones.
Applications to other systems and comparisons with experimental data are in progress.

\end{section}

\begin{acknowledgments}
We would like to acknowledge CNPq, FAPERJ BBP grant, FAPESP and the joint PRONEX initiatives of CNPq/FAPERJ under
Contract No.\ 26-111.443/2010, for partial financial support.
This work was supported in part by the National Science Foundation under Grant Nos. PHY-0606007 and INT-0228058.
\end{acknowledgments}

\bibliography{manuscript}
\bibliographystyle{apsrev4-1}

\end{document}